\documentclass[11pt]{elsarticle}
\usepackage{natbib}
\usepackage{stmaryrd}
\usepackage{subfigure}
\usepackage{color}
\usepackage{amsmath}
\usepackage{bm}
\usepackage{bbm}
\usepackage{mathrsfs}
\usepackage{graphicx}
\usepackage{wrapfig}
\usepackage{caption}
\usepackage{upgreek}
\usepackage{enumitem}
\usepackage{epsfig}
\usepackage{amsfonts}
\usepackage{amssymb}
\usepackage{amsmath}
\usepackage{wrapfig}
\usepackage{graphicx}
\usepackage{psfrag}
\usepackage{placeins}
\graphicspath{{./newFigures/}}
\def\bn{\mbox{\boldmath$ n$}}

\def\br{\mbox{\boldmath$ r$}}

\newtheorem{definition}{Definition}
\begin{document}
\title{A comparison of Redlich-Kister polynomial and cubic spline representations of the chemical potential in phase field computations}
\author[umme]{Gregory H. Teichert}
\author[ucsb]{N. S. Harsha Gunda}
\author[umme]{Shiva Rudraraju}
\author[ucsb]{Anirudh Raju Natarajan}
\author[ummse]{Brian Puchala}
\author[umme,umm]{Krishna Garikipati\corref{mycorrespondingauthor}}
\ead{krishna@umich.edu}
\author[ucsb]{Anton Van der Ven\corref{mycorrespondingauthor}}
\ead{avdv@engineering.ucsb.edu}
\cortext[mycorrespondingauthor]{Corresponding Authors}

\address[umme]{Mechanical Engineering Department, University of Michigan}
\address[ucsb]{Materials Department, University of California, Santa Barbara}
\address[ummse]{Materials Science and Engineering Department, University of Michigan}
\address[umm]{Mathematics Department, University of Michigan}

\begin{abstract}
Free energies play a central role in many descriptions of equilibrium and non-equilibrium properties of solids. 
Continuum partial differential equations (PDEs) of atomic transport, phase transformations and mechanics often rely on first and second derivatives of a free energy function. 
The stability, accuracy and robustness of numerical methods to solve these PDEs are sensitive to the particular functional representations of the free energy. 
In this communication we investigate the influence of different representations of thermodynamic data on phase field computations of diffusion and two-phase reactions in the solid state. 
First-principles statistical mechanics methods were used to generate realistic free energy data for HCP titanium with interstitially dissolved oxygen. 
While Redlich-Kister polynomials have formed the mainstay of thermodynamic descriptions of multi-component solids, they require high order terms to fit oscillations in chemical potentials around phase transitions.
Here, we demonstrate that high fidelity fits to rapidly fluctuating free energy functions are obtained with spline functions. Spline functions that are many degrees lower than Redlich-Kister polynomials provide equal or superior fits to chemical potential data and, when used in phase field computations, result in solution times approaching an order of magnitude speed up relative to the use of Redlich-Kister polynomials.

\end{abstract}


 \maketitle
\section{Introduction}

A free energy potential obtained by applying appropriate Legendre transforms to the internal energy encapsulates all the thermodynamic information that can be known about a bulk solid. Its first derivatives with respect to experimentally controlled state variables yield the equilibrium values of conjugate state variables that are not controlled experimentally. Its second derivatives are related to response functions such as heat capacities, compressibilities, elastic moduli etc. A free energy potential also codifies information about phase stability, and its rate of decrease points the direction of non-equilibrium processes in kinetic theories of phase transformations and microstructure evolution. The CALPHAD approach from its inception recognized the central role that free energies should play to ensure that experimental and calculated data is collected and organized in a thermodynamically self-consistent way \citep{Hillert2007,Kaufman2014}. 

The Gibbs and Helmholtz free energies are especially important in the study of multi-component solids. The Gibbs free energy is the potential used to calculate temperature composition phase diagrams at ambient pressure when assuming incoherent multi-phase coexistence \citep{Hillert2007}. The Helmholtz free energy is more appropriate when analyzing coherent multi-phase equilibrium, where the pressure is no longer uniform throughout the solid \citep{Voorhees2004}. 

A direct experimental measurement of a free energy is often not possible. Instead, free energies must be determined indirectly by integrating over measurable state variables. Chemical potentials, for example, can be determined electrochemically or by measurements of partial pressures. The integration of this data with respect to composition then yields a free energy \citep{DeHoff1993}. Free energies can also be calculated using first-principles statistical mechanics approaches that rely on effective Hamiltonians to extrapolate computationally expensive electronic structure calculations within Monte Carlo simulations \citep{Sanchez1984,deFontaine1994,Laks1992,Ceder1993,VanderVen1998,vandewalle2002a,vandewalle2002b,Arroyave2004,VanderVen2010,Thomas2013,Puchala2013,Chen2015,Natarajan2016}. While such approaches are rarely quantitatively accurate, they yield free energy descriptions that are physics based and that rigorously account for vibrational and configurational entropy.  

Independent of how a free energy description has been obtained there is a need to represent it mathematically. The CALPHAD approach maps the composition dependence of free energies on a polynomial expansion. A commonly used polynomial expansion is that introduced by Redlich and Kister \citep{RedlichKister1948}. While other polynomial expansions can also be used, ``mathematically, the choice of basis (for a finite-dimensional space) makes no difference'' according to Dahlquist and Bj\"{o}rck \citep{DahlquistBjorck2008}. It is known that sampling data at Chebyshev points significantly improves a polynomial {\it interpolation}, where the function passes through every data point. However, fitting to measured or calculated free energies generally involves many data points, making a least squares fit more appropriate than a polynomial interpolation. Chebyshev points are, therefore, not necessary \citep{DahlquistBjorck2008}. Because of the mathematical equivalence of the various polynomial expansions (Redlich-Kister expansion, Legendre polynomial series, simple power series, etc.) \citep{Tomiska1984,Ouerfellietal2010}, the effectiveness of a polynomial expansion can be assessed by considering one basis set. 
Even so, \citet{DahlquistBjorck2008} point out that some functions are ``not at all suited for approximation by one polynomial over the entire interval. One would get a much better result using approximation with piecewise polynomials.''

In this study, we compare the use of the Redlich-Kister polynomials with cubic splines (piecewise cubic polynomials with $C^2$ global continuity; i.e. continuous second derivatives) in fitting free energy data. This is done in the context of phase field modeling using the Cahn-Hilliard equation \citep{Cahn1958}. One physical phenomenon that this model captures is spinodal decomposition, where a material separates into two distinct phases. Spinodal decomposition arises when the free energy is concave with respect to composition resulting in a negative thermodynamic factor that causes uphill diffusion. Capturing this physics in a phase field model requires an accurate representation of the free energy and its higher order derivatives. We find that there are cases where even low-order splines are much more effective at representing the physics of the problem than are global polynomials, especially within the spinodal regions. We also see that the high polynomial degree sometimes required by the Redlich-Kister expansions significantly increases computation time.

The thermodynamics of our model, the titanium-oxygen system is discussed in Section \ref{sec:thermokin}. A detailed treatment of the different functional representations of the chemical potential and the free energy follows in Section \ref{sec:freeenergyreps}, supported by fits to a data set generated from first principles. Phase field computations is the subject of Section \ref{sec:phasefield}, where we focus on the consequences of the different chemical potential and free energy representations in terms of the physics and numerical performance. We place our results in perspective and make concluding remarks in Section \ref{sec:concl}.
 
\section{Model system: oxygen dissolved in HCP Ti}
\label{sec:thermokin}

We use the Ti-O binary as a model system to explore different representations of the composition dependence of free  energies. In contrast to many common metals, HCP titanium is capable of dissolving a high concentration of oxygen before other Ti suboxides having very different crystal structures form \citep{Cancarevic2007,Okamoto2011}. The dissolved oxygen fills octahedral interstitial sites of the HCP Ti crystal structure to form TiO$_x$ where $x$ measures the fraction of filled octahedral sites \citep{Burton2012}. The oxygen solubility limit in HCP Ti is as high as $x$=1/2  \citep{Cancarevic2007,Okamoto2011}. 

Since oxygen dissolves interestitially in HCP Ti, its chemical potential is related to the slope of the free energy according to 
\begin{align}
\mu_{O} = \left(\frac{\partial G}{\partial N_{O}}\right)_{N_{Ti}, T, P}=\left(\frac{\partial g}{\partial x}\right)_{T,P}
\label{eqn:mu_O}
\end{align}
provided that the total free energy of the solid, $G$, is normalized by the number of Ti atoms, $N_{Ti}$, with $g=G/N_{Ti}$ and $x=N_{O}/N_{Ti}$. 
A measurement of the oxygen chemical potential as a function of oxygen concentration $x$, enables a determination of the free energy through integration of Eq. (\ref{eqn:mu_O}).

\begin{figure}[tb!]
        \centering
        \includegraphics[width=\textwidth]{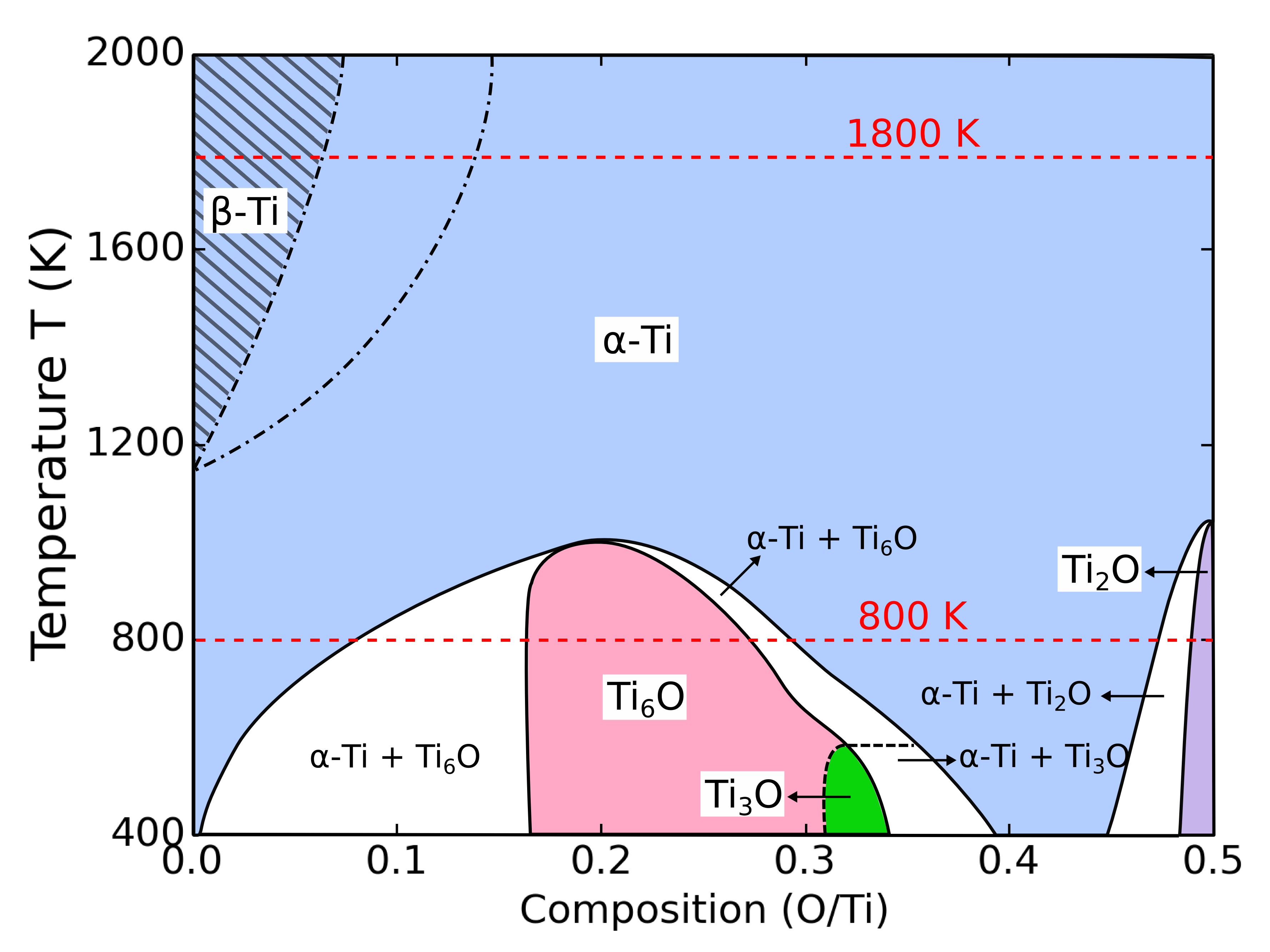}
        	\caption{The calculated temperature versus composition phase diagram of HCP TiO$_{x}$}
	\label{fig:phase_diagram}
\end{figure}

\begin{figure}[tb!]
        \centering
        \includegraphics[width=\textwidth]{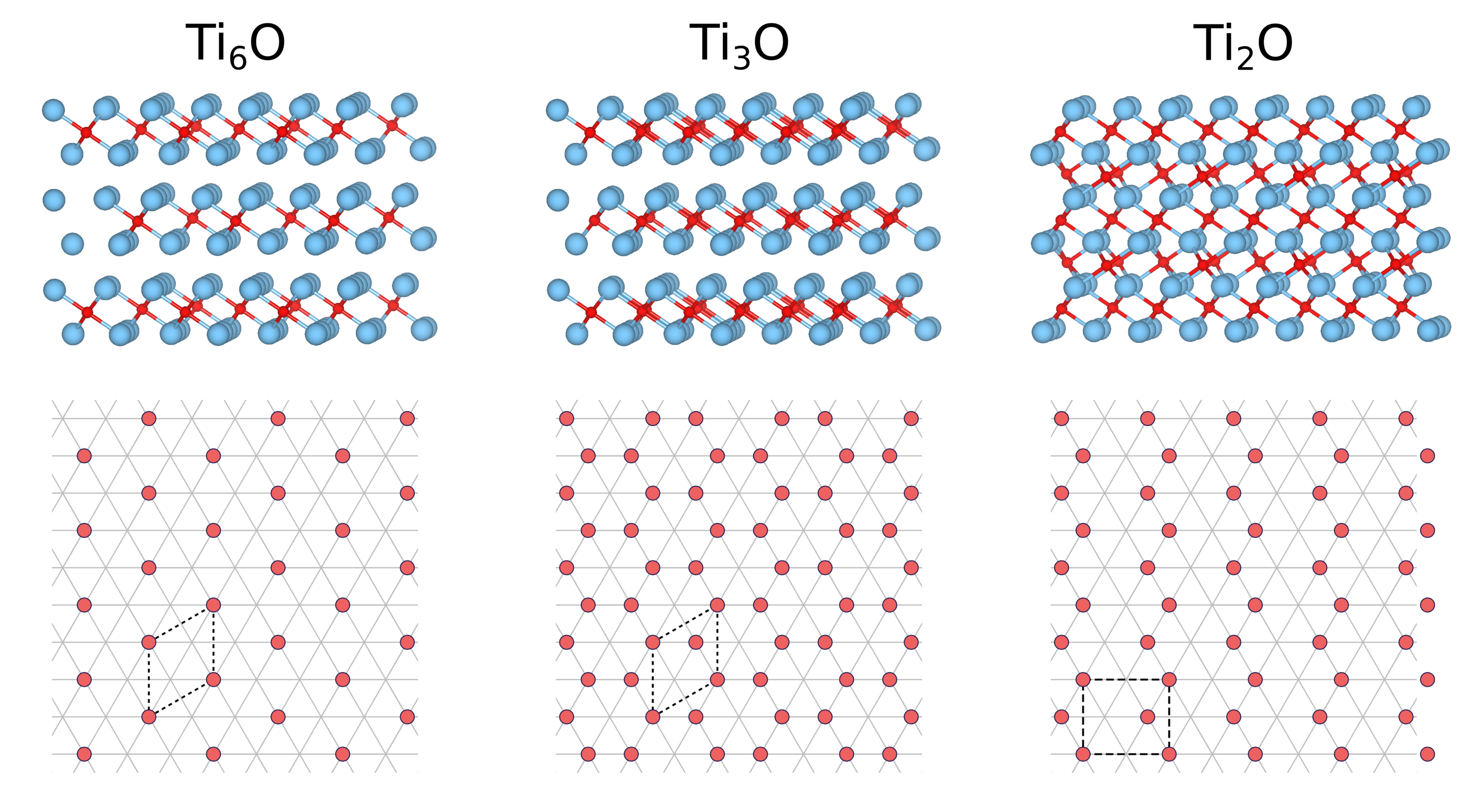}
        	\caption{The dissolved oxygen fills octahedral interstitial sites of the HCP Ti crystal structure to form TiO$_x$. Ordered phases are stable at low temperature at stoichiometric compositions of $x$=1/6, $x$=1/3 and $x$=1/2.}
	\label{fig:oxide}
\end{figure}

An important source of entropy in HCP TiO$_{x}$ arises from configurational disorder among oxygen and vacant interstitial sites. A first-principles statistical mechanics approach based on the cluster expansion is well suited to predict the composition dependence of its free energy\citep{Burton2012}. We used the CASM software package \citep{VanderVen2010,Thomas2013,Puchala2013} to calculate a first-principles phase diagram with a cluster expansion \citep{Sanchez1984,deFontaine1994} parameterized with density functional theory (DFT) energies and Monte Carlo simulations. The VASP plane-wave DFT code \citep{Kresse1996,Kresse1999} was used to calculate the energies of many different oxygen-vacancy orderings within HCP Ti. Details of these calculations and the parameterization of the cluster expansion will be published elsewhere. 

The calculated phase diagram of Figure \ref{fig:phase_diagram} shows that ordered phases are stable at low temperature at stoichiometric compositions of $x$=1/6, $x$=1/3 and $x$=1/2. The ordered phases are shown in Figure \ref{fig:oxide}. These ordered phases disorder upon heating to form a solid solution at high temperature. The octahedral interstitial sites of HCP form two-dimensional triangular lattices that are stacked directly on top of each other along the HCP $c$ axis. The ordered phases at $x$=1/6 and $x$=1/3 have a $\sqrt{3}a\times\sqrt{3}a$ supercell on the triangular lattices of octahedral sites within basal planes of HCP as illustrated in Figure \ref{fig:oxide}. Both ordered phases are staged in the sense that every other layer of interstitial octahedral sites is empty. Oxygen in the $x$=1/2 ordered phase arrange in a zig-zag pattern separated by a zig-zag pattern of vacant sites. The ordered phase at $x$=1/6 disorders around 1000K when at its stoichiometric composition while the ordered phase at $x$=1/2 disorders around 1100K. The ordered phases have a wide stability range along the composition axis due to their ability to tolerate anti-site defects. The ordered phase at $x$=1/6, for example, is predicted to remain thermodynamically stable to concentrations as high as $x$=0.25 at 600 K. The excess oxygen in the $x$=1/6 ordered phase is accommodated in a disordered fashion on the vacant sites within the filled layers. A second order phase transition separates the stability domain of the $x$=1/6 and $x$=1/3 orderings. The calculated phase diagram of Figure \ref{fig:phase_diagram} is qualitatively very similar to that predicted by Burton and van de Walle \citep{Burton2012} using a similar approach, although lower order-disorder transition temperatures are predicted in Figure \ref{fig:phase_diagram}. The phase diagram in Figure \ref{fig:phase_diagram} also shows the stability range of the BCC based $\beta$-TiO$_{x}$ phase observed experimentally \citep{Cancarevic2007,Okamoto2011}.

\begin{figure}[tb!]
        \centering
        \includegraphics[width=\textwidth]{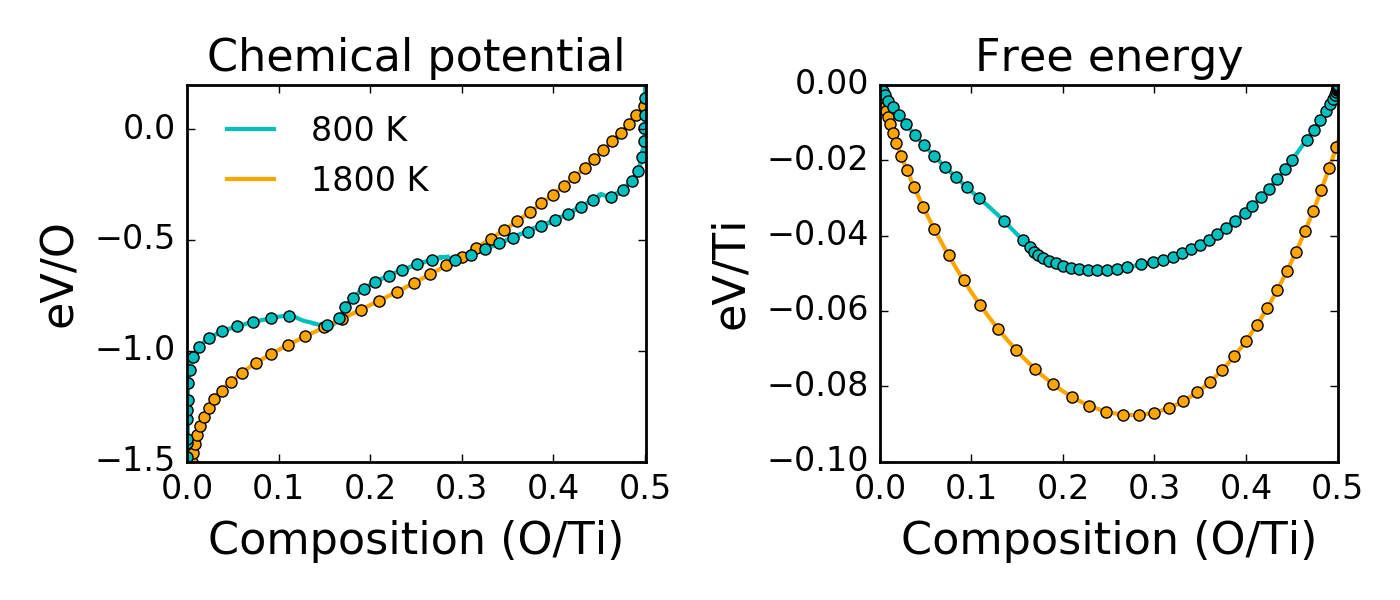}
        	\caption{The calculated oxygen chemical potentials and free energies at 1800 K and 800 K}
	\label{fig:free_energy}
\end{figure}

Figure \ref{fig:free_energy} shows the calculated oxygen chemical potentials and free energies for HCP TiO$_{x}$ at 1800 K and 800 K. The free energies were calculated by integrating the chemical potential versus the oxygen concentration \citep{Dalton2012}. At 1800 K, the disordered solid solution is stable between $x$=0 and $x$=1/2 as reflected by the sloping oxygen chemical potential and the convex nature of the free energy at this temperature. At 800 K, both the $x$=1/6 and $x$=1/2 ordered phases are stable with two-phase regions separating the ordered phases from the solid solution. The free energy within the two-phase regions is continuous due to the group/subgroup symmetry relation between the ordered phases and the disordered solid solution. We calculated the chemical potential within the two-phase region using variance constrained Monte Carlo simulations \citep{Sadigh2012}, which permits access to compositions where the free energy is concave and the solid is susceptible to decomposition.  

The ordered phases at $x$=1/6 and $x$=1/2 have a lower symmetry than the disordered solid solution TiO$_{x}$. Order parameters, $\vec{\eta}$, therefore exist that can distinguish the ordered phases from the disordered solid solution. Usually order parameters are defined such that they are zero in the disordered state ($\vec{\eta}=0$) and have finite values in the ordered phase (i.e. $\vec{\eta}=\vec{\eta}_{\alpha}$ for ordered phase $\alpha$). The free energy of the solid can therefore be expressed not only as a function of concentration, but also as a function of order parameters $g\left(x,\vec{\eta}\right)$. When a solid solution is thermodynamically stable, this free energy will exibit a minimum at $\vec{\eta}=0$. Likewise, at concentrations where a particular ordered phase $\alpha$ is stable, the free energy will have a minimum in the vicinity of $\vec{\eta}_{\alpha}$. The free energies depicted in Figure \ref{fig:free_energy}  can be viewed as the minimum of $g\left(x,\vec{\eta}\right)$ with respect to order parameters $\vec{\eta}$ at each concentration $x$, i.e. 
\begin{align}
g\left(x\right) = \min_{\vec{\eta}} g\left(x,\vec{\eta}\right)
\end{align}

\section{Different free energy representations}
\label{sec:freeenergyreps}
We compare the ability of a truncated Redlich-Kister series and a spline fit to faithfully represent the free energies of TiO$_{x}$.   We first consider the high temperature free energy curve corresponding to a solid solution. The low temperature free energy poses more challenges due to the presence of several two-phase regions where the free energy is concave. 

\subsection{Methodology}

Following standard CALPHAD practice, we express the free energy as 
\begin{align}
g(x) &= k_BT\left[x\log (x) + (1 - x)\log(1 - x)\right] + \Delta g(x)
\label{eqn:g(x)}
\end{align}
where the first term corresponds to an ideal solution entropy and $\Delta g(x)$ is an excess free energy representing a deviation from thermodynamic ideality. As usual $k_B$ is Boltzmann's constant and natural logarithms are used. The inclusion of an ideal solution entropy term is especially useful in describing the free energy in the dilute limits (i.e. $x \to 0$ and $x \to 1$) where alloys behave as ideal solutions and where the excess free energy goes to zero. 
With the above free energy expression, the oxygen chemical potential in TiO$_{x}$ then becomes 
\begin{align}
\mu(x) &= k_BT\log\left(\frac{x}{1-x}\right) + \Delta \mu(x)
\label{eqn:mu}
\end{align}
where $\Delta \mu(x)$ is the derivative with respect to $x$ of $\Delta g(x)$. For a substitutional binary A-B alloy, the chemical potential in Eq (\ref{eqn:mu}) corresponds to the difference in chemical potentials between A and B (i.e. $\mu=\mu_{B}-\mu_{A}$ if $x$ measures the concentration of B). The ideal solution term captures the logarthmic divergences in the chemical potential $\mu$ in the dilute limits ($x \to 0$ and $x \to 1$).

Below we compare the ability of a Redlich-Kister polynomial series and splines in fitting calculated values for the excess free energy $\Delta g$. Since the free energy is often found by first measuring or computing the chemical potential and then integrating, we fit $\Delta \mu$ to the difference between the chemical potential data and the logarithmic term $k_BT\log(x/(1-x))$.

\subsubsection{Redlich-Kister polynomial series}
We consider a R-K polynomial expansion for the function $\Delta g(x)$ of the following form:
\begin{align}
\Delta g(x) = g_0 + g_1x + x(1-x)\sum \limits_{k=0}^n L_k (2x-1)^k
\end{align}
Then the function $\Delta \mu$ used in the curve fitting of the chemical potential has the form:
\begin{align}
\Delta \mu = g_1 + \sum \limits_{k=0}^n L_k [2kx(1-x) - (2x-1)^2] (2x-1)^{k-1}
\label{eqn:RKmu}
\end{align}
The coefficients $L_k$ can be found by a least squares method.

We used the chemical potential data for the titanium oxide system for compositions from $x = 0$ to $x=\tfrac{1}{2}$.
To avoid the noise where the data is steepest near $x = 0$ and $x=\tfrac{1}{2}$, we only fit to the data between 0.001 and 0.499.

\subsubsection{Splines}
 
A spline is a piecewise polynomial with a specified order of continuity at the subdomain junctions. The endpoints of each subdomain are called \textit{knots} or \textit{breaks}. Quoting verbatim from \citet{DahlquistBjorck2008}, a more formal definition is as follows:

\begin{definition}
A spline function $s(x)$ of order $k \geq 1$ (degree $k-1 \geq 0$), on a grid
$$\Delta = \{a=x_0<x_1<\cdots<x_m=b\}$$
of distinct knots is a real function $s$ with the following properties:
\begin{enumerate}[label=(\alph*)]
\item For $x \in [x_i,x_{i+1}]$,$i=0:m-1$, $s(x)$ is a polynomial of degree $<k$.
\item For $k=1$, $s(x)$ is a piecewise constant function. For $k\geq2$, $s(x)$ and its first $k-2$ derivatives are continuous on $[a,b]$, i.e. $s(x) \in C^{k-2}[a,b]$
\end{enumerate}
\end{definition}

\begin{figure}[tb!]
        \centering
        \includegraphics[width=0.7\textwidth]{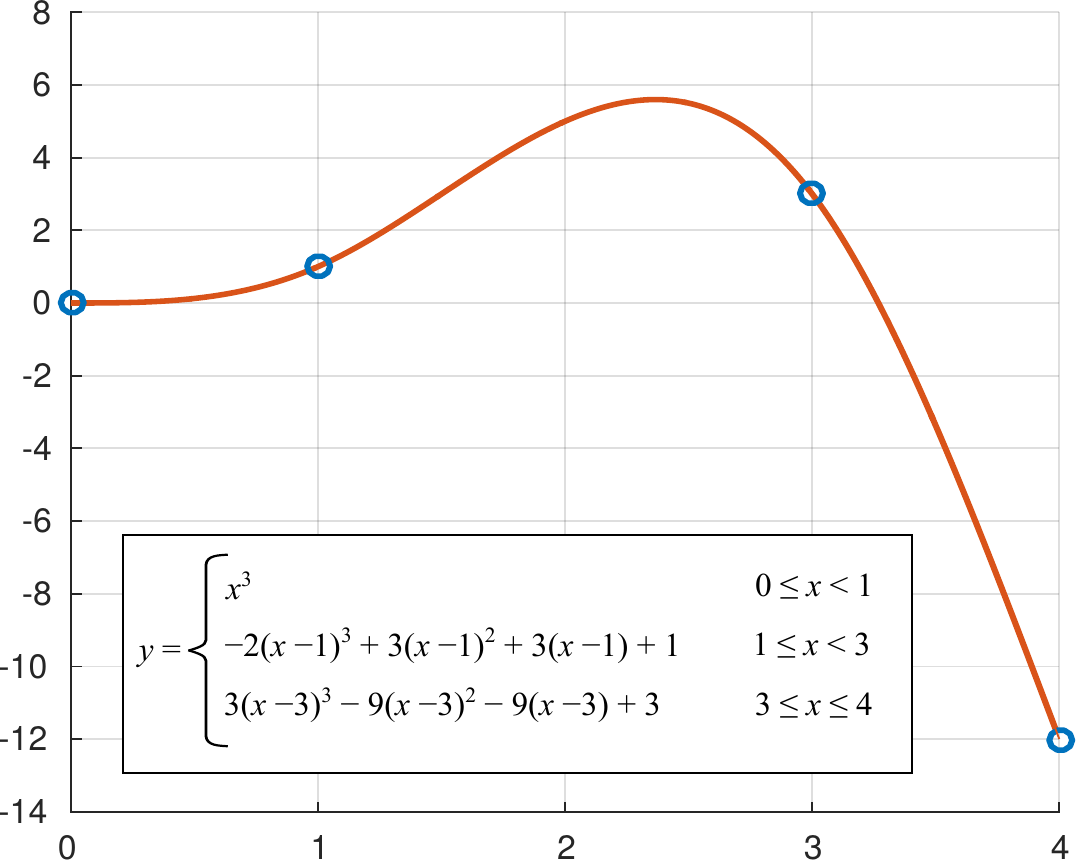}
        	\caption{An example cubic spline, defined as a piecewise cubic polynomial over three subdomains. The function values at the four knots are $\{(0,0),(1,3),(3,3),(4,-12)\}$. Note that the first and second derivatives are continuous at the knots.}
	\label{fig:spline_example}
\end{figure}

Note that the order of the spline is equal to the degree of the polynomial plus one. For example, cubic splines (order = 4) are piecewise cubic polynomials that are $C^2$ continuous across all knots (see Figure \ref{fig:spline_example}). Splines are able to capture local features  due to the local nature of piecewise polynomials. The continuity conditions at the knots maintain a specified global continuity across the entire function. These local and global characteristics of splines make them an important and often ideal tool in function interpolation and fitting.

There are multiple ways of representing splines. The simplest form is as a standard piecewise polynomial, where a standard polynomial is given for each subdomain. There is nothing in this structure that guarantees that the function is, in fact, a spline; the polynomial coefficients must be chosen to ensure the necessary continuity across knots. Splines may also be written as a linear combination of \textit{B-splines} (short for basis splines). We will denote a B-spline of order $k$ at knot $i$ by $N_{i,k}$, and we let $\tau_i$ be the location of knot $i$. Additional knots may be added at the endpoints, called \textit{exterior knots}. The original set of knots will be called \textit{interior knots}.  The B-spline $N_{i,k}$ is strictly positive on the interval $(\tau_i,\tau_{i+k})$ and zero outside the interval. Given $m+1$ interior knots, B-splines have the summation property $\sum_i N_{i,k}(x) = 1$, for $x \in [\tau_0,\tau_m]$ \citep{DahlquistBjorck2008}. B-splines can be represented using the following Cox-de Boor recurrence relation \citep{deBoor1972}:
\begin{align}
N_{i,k}(x) &= \frac{x - \tau_i}{\tau_{i+k-1}-\tau_i}N_{i,k-1}(x) + \frac{\tau_{i+k}-x}{\tau_{i+k}-\tau_{i+1}}N_{i+1,k-1}(x)\\
N_{i,1}(x) &= 
\begin{cases}
1 & \tau_i \leq x < \tau_{i+1}\\
0 & \text{otherwise}
\end{cases}
\end{align}
We can represent a spline function $s(x)$ of order $k$ with $m+1$ interior knots as a linear combination of B-splines:
\begin{align}
s(x) = \sum\limits_{i=-k+1}^{m-1}c_iN_{i,k}
\label{eqn:Bspline}
\end{align}
where $c_i$ are the appropriate B-spline coefficients. The piecewise polynomial representation is
\begin{equation}
\begin{aligned}
s(x) &= \sum\limits_{i=0}^{m-1}P_{i,k}\\
P_{i,k} &= 
\begin{cases}
\sum\limits_{j=0}^{k-1} D_{i,j}(x-\tau_i)^{k-1-j} & \tau_i \leq x < \tau_{i+1}\\
0 & \text{otherwise}
\end{cases}
\end{aligned}
\label{eqn:ppspline}
\end{equation}
where $\{D_{i,0},\ldots,D_{i,k-1}\}$ are the coefficients for the polynomial in subdomain $i$.

In the case of spline interpolation, each given data point becomes a knot in the spline function. For $n$ given data points, there are $n$ knots and $n-1$ subdomains. For a spline of order $k$, there are $k$ polynomial coefficients within each subdomain, for a total of $k(n-1)$ unknowns for the entire spline. The function value at each subdomain endpoint is specified, giving $2(n-1)$ constraints. This gives at  least $C^0$ continuity across the function. Specifying continuity of order $m$ at each subdomain junction adds $m(n-2)$ total constraints, for a total of $2(n-1) + m(n-2)$ constraints. Since the number of constraints must be less than or equal to the number of unknowns, we find that function interpolation using $k$\textsuperscript{th} order splines allows $C^{k-2}$. The interpolation is performed by solving for the unknown with these constraints. This interpolation process can be done using splines written as a linear combination of B-splines or using the simple piecewise polynomial representation.

Where many data points are given, as in this study, it is useful to perform a spline fit instead of a spline interpolation. This allows fewer knots in the spline than data points, which can smooth out noise in the data and simplify the evaluation of the function. The curve fit can be done using a least square spline approximation. The B-spline representation of a spline allows us to express the least squares problem as follows:
\begin{align}
\min\limits_{\vec{\mathbf{c}}}\sum\limits_{j=1}^n\left(\sum\limits_{i=-k+1}^{m-1}c_iN_{i,k}(x_j)-f_j\right)^2
\end{align}
where $\vec{\mathbf{c}}$ is the vector of coefficients, $k$ is the B-spline order, $m+1$ is the number of interior knots, and $(x_j,f_j)$ is the $j$\textsuperscript{th} data point \citep{DahlquistBjorck2008}.

We use cubic splines to fit $\Delta \mu$ in the chemical potential function. We perform a spline fit, as opposed to a spline interpolation, to smooth out the noise from the data and simplify the representation and evaluation of the function. The Octave \texttt{splinefit} function takes as input the data points and allows us to specify the location of the knots in the spline fit. It performs a least squares fit to the given data using B-splines. However, this function returns the spline's coefficients for the simple piecewise polynomial form shown in Eq. (\ref{eqn:ppspline}). This set of coefficients can then be evaluated for a given value of $x$ in Octave using the \texttt{ppval} function or manually as standard piecewise polynomials.

To accurately capture local features of the data while minimizing the number of knots, we first identify regions where the data are changing rapidly. In the final spline fit, we use a higher density of knots in those regions. To do this, we first fit a cubic spline to the data with knots at relatively small intervals of 0.005 to approximate $\Delta \mu$. We use this initial spline fit of $\Delta \mu$ to approximate $\Delta \mu''(x)$, and we identify any region where $|\Delta \mu''(x)| > 600$ as a region with rapidly changing data. For the final fit, we specify knots at intervals of 0.004 within the regions of rapidly changing data and at intervals of 0.03 outside those regions. This allows us to capture all the physics while reducing the number of knots.  The function \texttt{splinefit} is used with this new set of knots to fit the data and update our spline representation of $\Delta \mu$.

\subsection{Results}

\subsubsection{Data with no spinodal regions}

\begin{figure}[tb!]
        \centering
        \includegraphics[width=\textwidth]{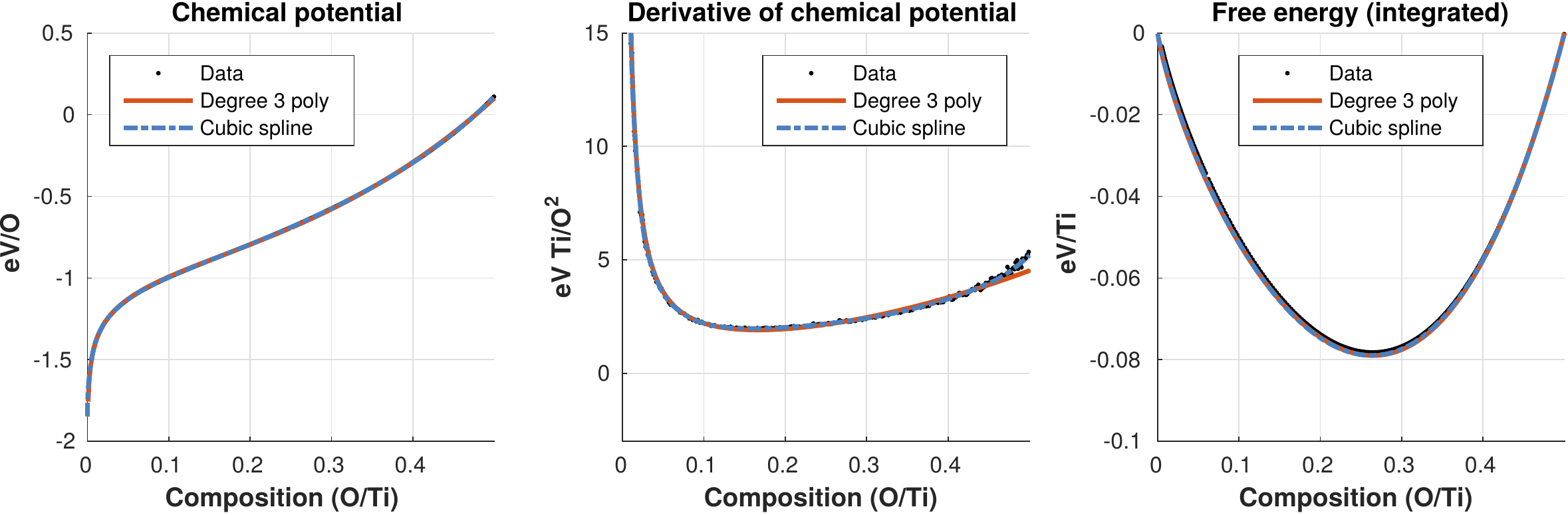}
        	\caption{A degree three R-K polynomial provides a sufficient fit to the chemical potential for titanium oxide at 1800 K, as does a cubic spline.}
	\label{fig:1800fit}
\end{figure}

At 1800 K, the titanium oxide system has a simple free energy landscape with no spinodals, and a degree three R-K polynomial expansion gives a sufficient fit of the chemical potential, as does the cubic spline. The chemical potential data and curve fit are shown in the left plot of Figure \ref{fig:1800fit}. The center plot shows the derivative of the chemical potential fit and the numerically differentiated data. This is also the second derivative of the free energy with respect to composition, which shows where the free energy curve is concave or convex. When the second derivative of the free energy is negative, the free energy is concave and represents a two-phase region. The numerical derivative was taken by first smoothing the data with a running average to remove some of the noise, then performing a central difference derivative. The plot on the right shows the free energy found by integrating the curve fit and data. It is plotted with respect to the end members at $x = 0$ and $x = \tfrac{1}{2}$. We also present a cubic spline fit of the data. Note that the cubic spline, while it also fits the data well, does not give a significantly better representation of this set of data.

\subsubsection{Data with three spinodal regions}

\begin{figure}[tb!]
        \centering
        \includegraphics[width=\textwidth]{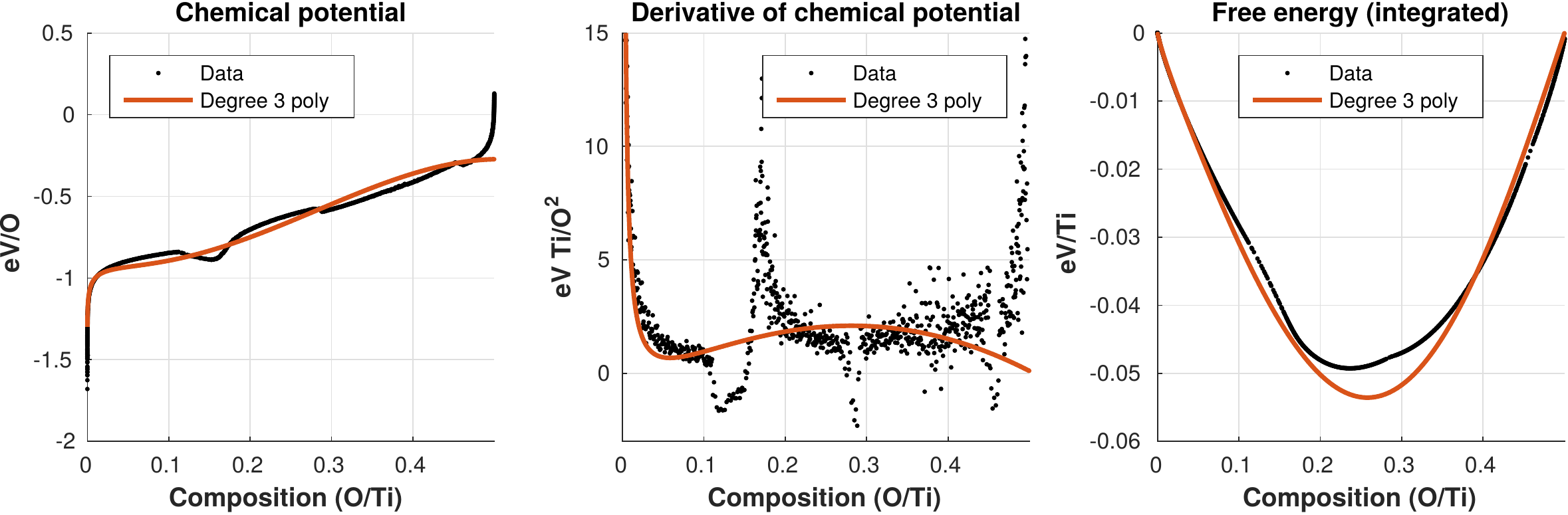}
        	\caption{A degree three R-K polynomial with the term $\log(x/(1-x))$ fails to capture the divergent behavior in the chemical potential at $x = \tfrac{1}{2}$ for titanium oxide at 800 K.}
	\label{fig:800_0to1}
\end{figure}

At 800 K, the HCP form of TiO$_{x}$ exhibits three spinodal regions. To accurately capture this physics, the function representing the free energy must have three corresponding concave regions. Fewer data points were calculated with the variance constrained Monte Carlo within these spinodal regions. They were weighted $10\times$ to help capture the spinodals with the polynomial function. The data are fitted with a degree three R-K polynomial as before. The data and fit are plotted in Figure \ref{fig:800_0to1} along with their corresponding derivatives and free energy. Note that the chemical potential data was smoothed before computing the numerical data to reduce the effects of noise. Also, as in all plots in this work, the free energies are plotted with respect to the end members at $x = 0$ and $x = \tfrac{1}{2}$.

\begin{figure}[tb!]
        \centering
        \includegraphics[width=\textwidth]{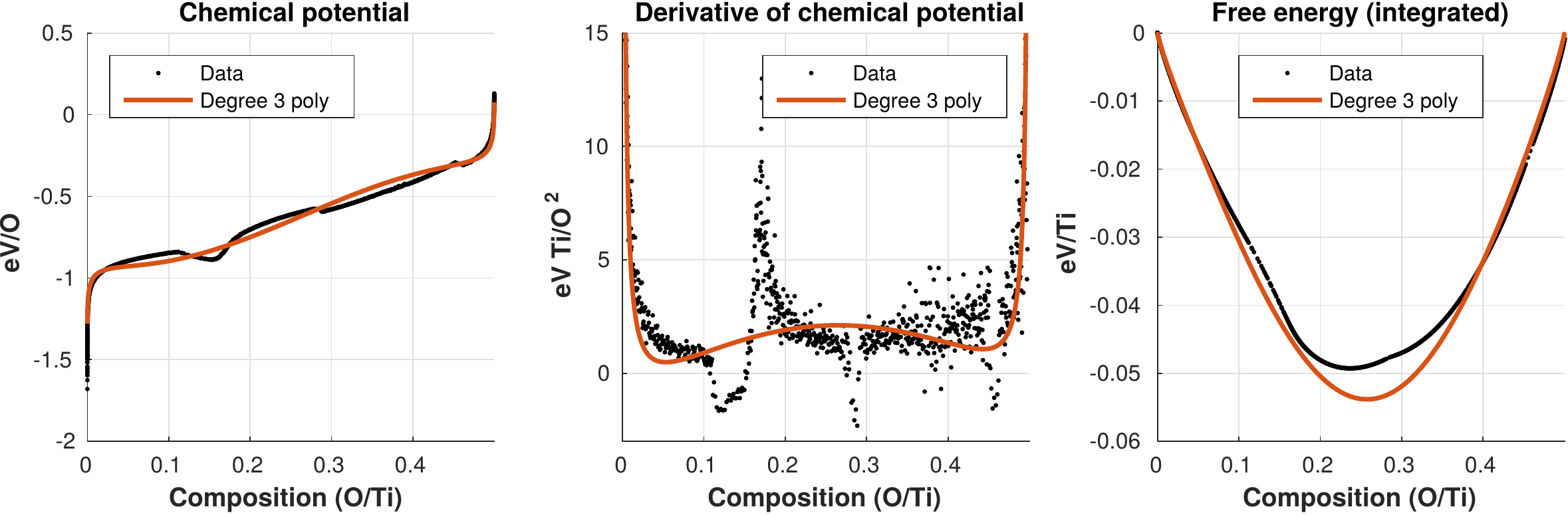}
        	\caption{The term $\log(2x/(1-2x))$ is able to capture the divergent behavior in the chemical potential for titanium oxide at 800 K.}
	\label{fig:800_0tohalf}
\end{figure}

The data diverge due to an ordering at $x = \tfrac{1}{2}$, but the equation used here does not capture this divergence.
We can represent the divergent behavior of the system by scaling the logarithmic term. This gives us the modified form
\begin{align}
\mu(x) = k_BT\log\left(\frac{2x}{1-2x}\right) + \Delta \mu(x)
\label{eqn:muScaled}
\end{align}
to model the chemical potential curve over the domain $\left[0,\tfrac{1}{2}\right]$. This corresponds to a free energy density of
\begin{align}
g(x) &= \tfrac{1}{2}k_BT\left[2x\log (2x) + (1 - 2x)\log(1 - 2x)\right] + \Delta g(x)
\label{eqn:gScaled}
\end{align}
For consistency, we also rescale the Redlich-Kister polynomial to the same domain of $\left[0,\tfrac{1}{2}\right]$.

\begin{figure}[tb!]
        \centering
        \includegraphics[width=\textwidth]{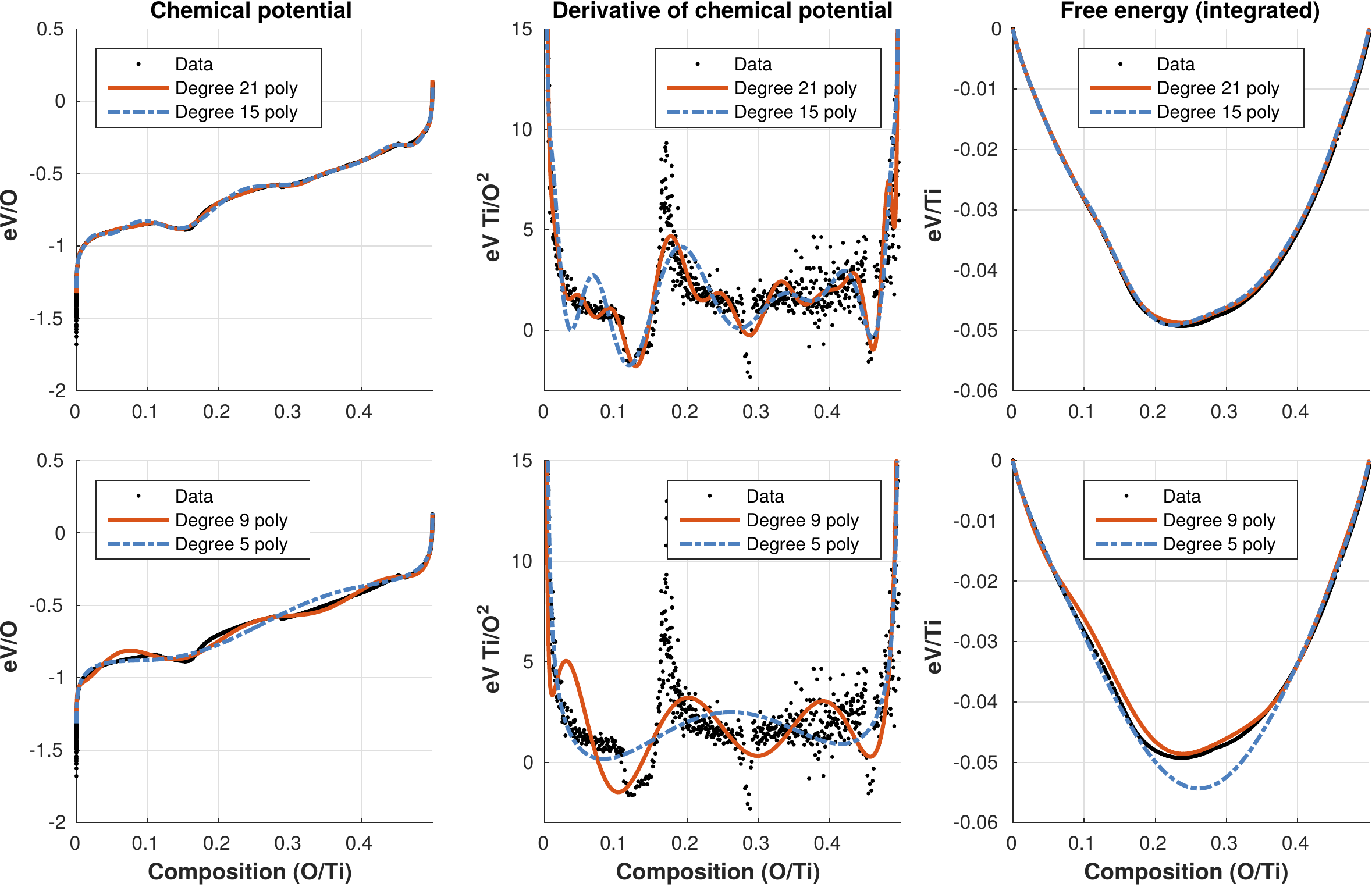}
        	\caption{R-K polynomials of degree 5, 9, 15, and 21 are used to fit the chemical potential data. The corresponding numerical and analytical derivatives are plotted. The free energy is found from the chemical potential integral and is with respect to end members at $x=0$ and $x=\tfrac{1}{2}$. Only the degree 21 polynomial captures all three spinodals.}
	\label{fig:polyFit}
\end{figure}

\begin{figure}[tb!]
        \centering
        \includegraphics[width=\textwidth]{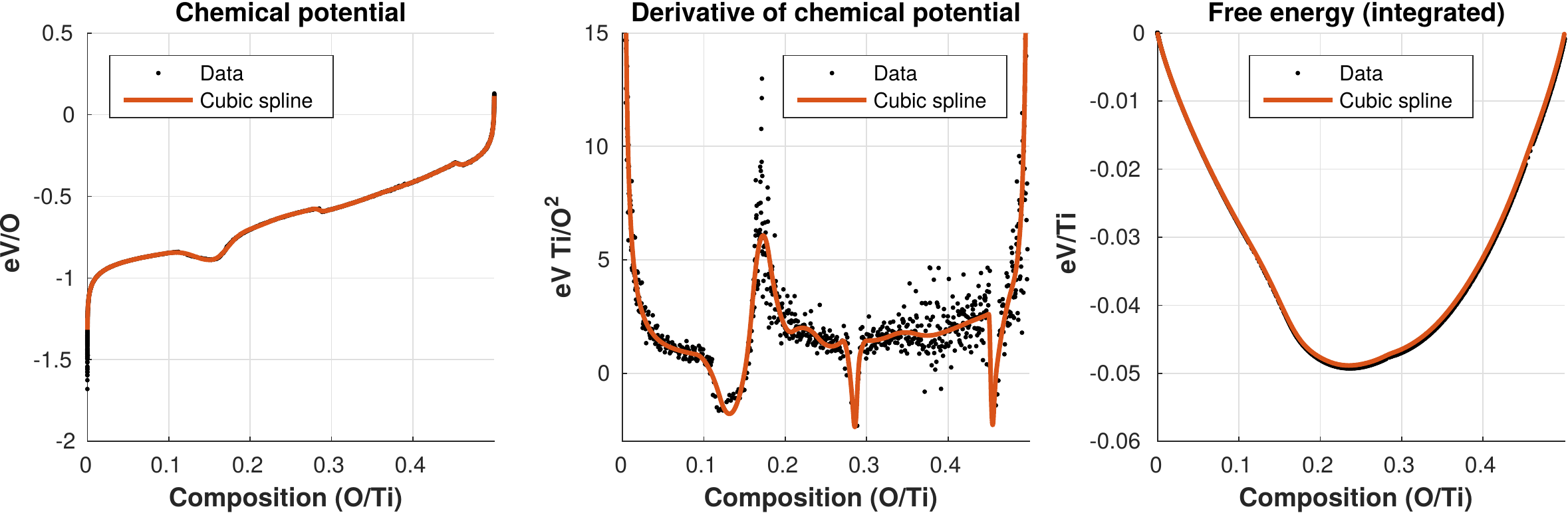}
        	\caption{The chemical potential data is fit using a cubic spline with 25 knots. The corresponding numerical and analytical derivatives are plotted. The free energy is found from the chemical potential integral and is with respect to end members at $x=0$ and $x=\tfrac{1}{2}$. The spline accurately represents all three spinodals.}
	\label{fig:splineFit}
\end{figure}

We now fit $\Delta \mu(x)$ to the difference between the chemical potential data and $k_BT\log(2x/(1-2x))$. The resulting fit is shown in Figure \ref{fig:800_0tohalf}. Note that the divergence of the data at $x = \tfrac{1}{2}$ is now represented in the curve fit. It is clear, however, that a degree three R-K polynomial cannot represent the spinodal regions. Polynomials of higher degree are used to fit the chemical potential and are plotted in Figure \ref{fig:polyFit}. The curve fit represents a spinodal region wherever the derivative of the chemical potential is negative, meaning the free energy is concave (see the center plots of Figure \ref{fig:polyFit}). Using the curve fit methods described here, the degree 21 polynomial capture all three spinodals. The degree 15 polynomial represents only two (the first and third), and it nearly produces a spurious spinodal at composition 0.05. The polynomial of degree nine captures one (the first spinodal), and degree five polynomial does not capture any spinodal regions. Clearly, to accurately represent all three spinodals with a R-K polynomial, a polynomial of high degree must be used.

A cubic spline was also used to fit $\Delta\mu$. Compare these results, shown in Figure \ref{fig:splineFit}, with the previous R-K polynomial fits. The spline clearly captures all three two-phase regions. This is done with minimal spurious oscillations in the derivative of the chemical potential and no spurious spinodal regions (see the center plot of Figure \ref{fig:splineFit}).

It should be noted that the failure of low order Redlich-Kister polynomials to capture the various spinodals of the first-principles free energy curve of TiO$_{x}$ at 800K signifies that the initial model based on Eq. (\ref{eqn:gScaled}) and Eq. (\ref{eqn:muScaled}) is too simple to capture the complex entropic contributions when ordered and disordered phases can coexist. 
The logarithmic composition dependence of Eq. (\ref{eqn:g(x)}) and Eq. (\ref{eqn:mu}) represents configurational entropic contributions of a thermodynamically ideal solid solution. 
Strong deviations from thermodynamic ideality occur at stoichiometric compositions where ordering occurs. 
This is evident in the oxygen chemical potential of TiO$_{x}$ at 800K shown in Figure 3 where steps appear at x=1/6 and 1/2. 
The non-ideal behavior at compositions where ordered phases are stable are more accurately captured with compound free energy models \cite{Hillert2001}. 
However to date, these models have been developed on a case by case basis and can become quite cumbersome for complex ordered phases such as the substoichiometric TiO$_{x}$ oxides at x=1/6, 1/3 and 1/2. 
They rely on sublattice concentration variables and approximate the entropy assuming random mixing within each sublattice. 
As such models are more accurate in describing entropy due to dilute mixing on the different sublattices of an ordered phase, they require less terms in Redlich Kister expansions that describe the excess free energy. 
While such an approach is desirable for a rigorous CALPHAD assessment of phase stability\cite{Fischer1997,Waldner1999}, it is less practical when developing free energy descriptions for phase field simulations due to the presence of additional sublattice composition variables \cite{Zhu2002,Kitashima2008,Zhang2015}. 
In that respect, splines, as will become evident in the next section are more versatile in empirically representing experimental or first-principles free energy curves needed as input for kinetic modeling.

\section{Phase field computations}
\label{sec:phasefield}

In this section we explore the extent to which phase-field simulations are affected by different representations of the free energy. To this end we perform phase field simulations of oxygen indiffusion in metallic Ti using the various Redlich-Kister polynomial parameterizations and the spline fits of the free energy of HCP TiO$_{x}$ described in the previous section. 

Oxygen indiffusion in HCP Ti is an important process in a variety of applications, the most obvious one being during the oxidation of Ti \citep{Unnam1986, Pouilleau1997}. The exposure of Ti to an oxygen rich environment quickly results in the formation of oxides such as rutile \citep{Ting2000,Krishna2005,Jamesh2013} or anatase TiO$_{2}$ and rock salt TiO \citep{Kao2011,Chung2011,Chung2012,Okazumi2011}. A layered microstructure typically forms with the most oxygen rich phases appearing at the surface and the oxides having a lower oxygen to titanium ratio forming below the surface \citep{Unnam1986}. The high oxygen solubility within HCP Ti should also result in a sizable region below these distinct oxide phases that is characterized by a large oxygen concentration gradient that drives further oxygen diffusion towards the interior of the metal \citep{Unnam1986, Dong1997,Dong2000}. While Ti-oxides with high oxygen content form at ambient oxygen partial pressures, they can be suppressed at sufficiently low oxygen partial pressures where they are no longer thermodynamically stable \citep{Dong2000}. Under those conditions, oxygen will simply dissolve in HCP Ti and form an HCP-based TiO$_{x}$ solid solution at the surface along with the ordered HCP TiO$_{x}$ phases having $x$=1/6, 1/3 and 1/2. 

Similar oxygen diffusion processes in HCP Ti are relevant within the recently proposed Ti-air battery, consisting of a Ti anode separated from an oxygen rich cathode by a solid-oxide electrolyte phase \citep{VanderVen2013a}. The externally imposed voltage (EMF) enables control over the oxygen chemical potential of the anode and high voltages and capacities are theoretically achievable when cycling the Ti anode between pure Ti and oxygen rich TiO$_x$ solid solutions ($x < 0.5$). By limiting the voltage to a narrow window, it is possible to suppress the formation of the oxygen rich Ti-oxides. 

Here we use phase field simulations to model oxygen diffusion in HCP Ti assuming zero flux boundary conditions on the surface. This approximates the process of oxygen diffusion through the metallic Ti substrate of an oxidizing sample and through the Ti anode during electrochemical cycling of a Ti-air battery after the source of oxygen at the surface has been cut off. 

The TiO$_{x}$ solid solution as well as the ordered TiO$_{1/6}$, TiO$_{1/3}$ and TiO$_{1/2}$ suboxides all share the same HCP Ti sublattice. The ordered suboxides therefore have a group/subgroup symmetry relationship with the disordered TiO$_{x}$ solid solution. This property makes it possible to describe oxygen indiffusion and order-disorder phase transformations with a combined Cahn-Hilliard and Allen-Cahn description \citep{Cahn1958,Hilliard1970,Allen1979,Chen2002,Balluffi2005}. The overall free energy of the solid can be written as an integral over a free energy density that is a function of the local concentration $x(\vec{\br})$ and local values of the order parameters $\vec{\eta}(\vec{\br})=\left(\eta_{1}(\vec{\br}),...,\eta_{j}(\vec{\br}),...,\eta_{n}(\vec{\br})\right)$ as well as the gradients of the local composition and order parameters according to:

\begin{align}
{\textbf F} = \int_{V} \frac{d\vec{\textbf r}}{\Omega} \left[g\left(x(\vec{\textbf r}), \vec{\eta}(\vec{\textbf r})\right)+\sum_{\alpha,\beta}\kappa_{\alpha,\beta}\frac{\partial x}{\partial r_{\alpha}}\frac{\partial x}{\partial r_{\beta}} +\sum_{i,j}\sum_{\alpha,\beta}\Gamma_{\alpha,\beta}^{i,j}\frac{\partial \eta_{i}}{\partial r_{\alpha}}\frac{\partial \eta_{j}}{\partial r_{\beta}}\right]
\label{eqn:total free energy}
\end{align}
where $\alpha$ and $\beta$ correspond to the Cartesian $x$, $y$ and $z$ directions and where $\Omega$ is the volume of TiO$_{x}$ per Ti atom. The tensor elements, $\kappa_{\alpha,\beta}$ and $\Gamma_{\alpha,\beta}^{i,j}$, represent gradient energy coefficients. In the above free energy expression, we have neglected gradient energy terms that couple a gradient in concentration $x$ and a gradient in the order parameters $\eta_{j}$, which may be allowed depending on how the order parameters transform under symmetry operations of the crystal. 

The temporal evolution of the solid when the initial total free energy is not minimal can be described with coupled Cahn-Hilliard and Allen-Cahn evolution equations. The Cahn-Hilliard equation in this example describes long-range oxygen diffusion with the oxygen flux, $\vec{J}$, driven by gradients in the oxygen chemical potential according to \citep{deGroot1984}
\begin{align}
\vec{J}=-L\nabla \mu
\label{eqn:flux}
\end{align}
where $L$ is the Onsager transport coefficient \citep{VanderVen2010b,VanderVen2013b} for oxygen diffusion over the interstitial sites of HCP Ti. Combining this flux expression with the continuity equation (i.e. conservation of oxygen atoms)
\begin{align}
\frac{\partial c}{\partial t}=-\nabla{\vec{J}}
\label{eqn:continuity}
\end{align}
yields a Cahn-Hilliard differential equation that links the time derivative of the oxygen concentration $c=x/\Omega$ to spatial derivatives of the oxygen chemical potential $\mu$. The oxygen chemical potential in a solid having a non-uniform concentration profile is the variational derivative of the total free energy
\begin{align}
\mu=\frac{\delta \textbf{F}}{\delta x}
\label{eqn:mu}
\end{align}
These equations are sufficient to describe oxygen redistribution in the solid solution phase above the order-disorder transition temperatures of TiO$_{1/6}$, TiO$_{1/3}$ and TiO$_{1/2}$. However, at 800 K, a concentration gradient will result in local concentrations where the ordered TiO$_{1/6}$ and TiO$_{1/2}$ suboxides are stable. An Allen-Cahn description is then necessary to complement the above Cahn-Hilliard equations to describe oxygen ordering tendencies. For each order parameter $\eta_{i}$, there is an Allen-Cahn evolution equation of the form \citep{Chen2002}
\begin{align}
\frac{\partial \eta_{i}}{\partial t}=-\sum_{j} M_{ij}\frac{\delta \textbf{F}}{\delta \eta_{j}}
\label{eqn:Allen-Cahn}
\end{align}
where M$_{ij}$ is a kinetic coefficient that determines the rate of change in the degree of local ordering in the presence of a free energy driving force. It is determined by local atomic hop events that are required to change the local degree of ordering. 

Since ordering processes at fixed concentration require only short-range diffusion involving a small number atomic hops, they will relax substantially more rapidly than processes that require long-range diffusion as described by the Onsager transport coefficients, $L$. Hence, while both Cahn-Hilliard and Allen-Cahn processes occur simultaneously, the Allen-Cahn evolution equations will reach a local minimum much more rapidly than the long-range diffusion processes described by Cahn-Hilliard. This therefore motivates an effective Cahn-Hilliard dynamics, in which Allen-Cahn ordering processes occur sufficiently rapidly that the local order parameters, $\vec{\eta}(\vec{\textbf{r}})$, are equal to the values that minimize the free energy at the local concentration $x(\vec{\textbf{r}})$. This has as a consequence that the order parameters are functions of the local concentration, i.e. $\vec{\eta}(x(\vec{\textbf{r}}))$. The above total free energy, Eq. (\ref{eqn:total free energy}), can then be converted into a free energy that only explicitly depends on the concentration profile
\begin{align}
{\textbf F} = \int_{V} \frac{d\vec{\textbf r}}{\Omega} \left[g\left(x(\vec{\textbf r})\right)+\sum_{\alpha,\beta}\tilde{\kappa}_{\alpha,\beta}\frac{\partial x}{\partial r_{\alpha}}\frac{\partial x}{\partial r_{\beta}}\right]
\label{eqn:effective total free energy}
\end{align}
where the homogeneous free energy $g$ is defined according to Eq. (\ref{eqn:gScaled}) and the gradient energy coefficients $\tilde{\kappa}_{\alpha,\beta}$ are functions of $\kappa_{\alpha,\beta}$, $\Gamma_{\alpha,\beta}^{i,j}$ and the equilibrium relationships between order parameters and composition $\vec{\eta}(x)$.

The Cahn-Hilliard phase field simulations of oxygen diffusion in HCP Ti implicitly rely on the above assumptions. We also neglect contributions to the free energy from coherency strains due to the dependence of lattice parameters on composition. These effects can be accounted for by explicitly making the free energy expression a function of local strains and variationally minimizing with respect to a displacement field to yield mechanical equilibrium criteria \citep{Chen2002,Voorhees2004,Rudraraju2014,Rudraraju2016}. While the equilibrium volume of HCP based TiO$_{x}$ does depend on $x$, this variation is at most only several percent \citep{VanderVen2013a}. 

The resulting equation for the chemical potential, equal to the variational derivative of the free energy in Eq. (\ref{eqn:effective total free energy}), is given by
\begin{equation}
\begin{aligned}
\mu &= \frac{1}{\Omega} \left[\frac{\partial g}{\partial x} - 2\sum_{\alpha,\beta}\tilde{\kappa}_{\alpha,\beta}\frac{\partial^2 x}{\partial r_{\alpha}\partial r_{\beta}}\right]
\end{aligned}
\end{equation}
where $\partial g/\partial x = \bar{\mu}$ is the homogeneous chemical potential as defined in Eq. (\ref{eqn:mu_O}) and (\ref{eqn:muScaled}). Considering the case where $\tilde{\kappa}_{\alpha,\beta} = (\hat{\kappa}/2)\delta_{\alpha\beta}$ simplifies this equation further:
\begin{equation}
\mu = \frac{1}{\Omega} \left(\bar{\mu}  - \hat{\kappa}\nabla^2 x\right)
\end{equation}
By inserting this equation for the chemical potential into Eq. (\ref{eqn:flux}) and (\ref{eqn:continuity}), we have the following Cahn-Hilliard equation:
\begin{equation}
\begin{aligned}
\frac{\partial x}{\partial t} &=\nabla \cdot\left[ L\nabla\left(\bar{\mu} - \hat{\kappa}\nabla^2 x\right)\right]
\end{aligned}
\end{equation}
The parameter $\hat{\kappa}$ and the average homogeneous chemical potential give a length scale for the interface $\sim\sqrt{\hat{\kappa}/|\bar{\mu}_{avg}|}$. Substituting the expression for the homogeneous chemical potential in Eq. (\ref{eqn:muScaled}) gives the following:
\begin{align}
\frac{\partial x}{\partial t} &= \nabla \cdot\left[ L\nabla\left(k_BT\log\left(\frac{2x}{1-2x}\right) + \Delta \mu - \hat{\kappa}\nabla^2 x\right)\right]
\label{eqn:dxdt}
\end{align}
The corresponding boundary conditions are
\begin{equation}
\begin{aligned}
\nabla x\cdot\vec{\bn} &= 0,\\
L\nabla\mu\cdot\vec{\bn} &= \vec{J},
\end{aligned}
\label{eqn:bcs}
\end{equation}
where the first boundary condition results from assuming equilibrium at the boundary, and the second represents an influx $\vec{J}$ at the boundary.

Because the cubic spline representation of the excess chemical potential is central to this study, we insert the spline function into Eq. (\ref{eqn:dxdt}).  The final form for the local Cahn-Hilliard equation, using the piecewise polynomial spline representation from equation (\ref{eqn:ppspline}) for $\Delta\mu$, is given by
\begin{equation}
\frac{\partial x}{\partial t} = \nabla \cdot\left[ L\nabla\left(k_BT\log\left(\frac{2x}{1-2x}\right) +  \sum\limits_{i=0}^{m-1}P_{i,4} - \hat{\kappa}\nabla^2 x\right)\right]
\end{equation}
where the piecewise cubic polynomials $P_{i,4}(x)$ are given by
\begin{equation}
P_{i,4} = 
\begin{cases}
\sum\limits_{j=0}^{3} D_{i,j}(x-\tau_i)^{3-j} & \tau_i \leq x < \tau_{i+1}\\
0 & \text{otherwise}
\end{cases}
\end{equation}
The polynomial coefficients $D_{i,j}$ used in the computations presented here were obtained using the Octave \texttt{splinefit} function.

As described previously, the spline function can also be expressed as a linear combination of B-splines, following Eq. (\ref{eqn:Bspline}). This alternative form gives the following Cahn-Hilliard equation:
\begin{align}
\frac{\partial x}{\partial t} &= \nabla \cdot\left[ L\nabla\left(k_BT\log\left(\frac{2x}{1-2x}\right) +  \sum\limits_{i=-3}^{m-1}c_iN_{i,k}(x)) - \hat{\kappa}\nabla^2 x\right)\right]
\end{align}

In addition to the phase field computations using a cubic spline fit for the excess chemical potential, we also performed simulations using the Redlich-Kister polynomial representation. A comparison of the resulting composition profiles, total energy, and computation times is presented here.

We solve the weak form of the Cahn-Hilliard equation with Isogeometric Analysis (IGA), which is a mesh-based numerical method that uses NURBS (Non-Uniform Rational B-Splines) as basis functions \citep{Hughesetal2005,Cottrelletal2009}. The code was implemented using the \texttt{PetIGA} library \citep{Collieretal2013}.

\begin{figure}[tb!]
        \centering
        \includegraphics[width=\textwidth]{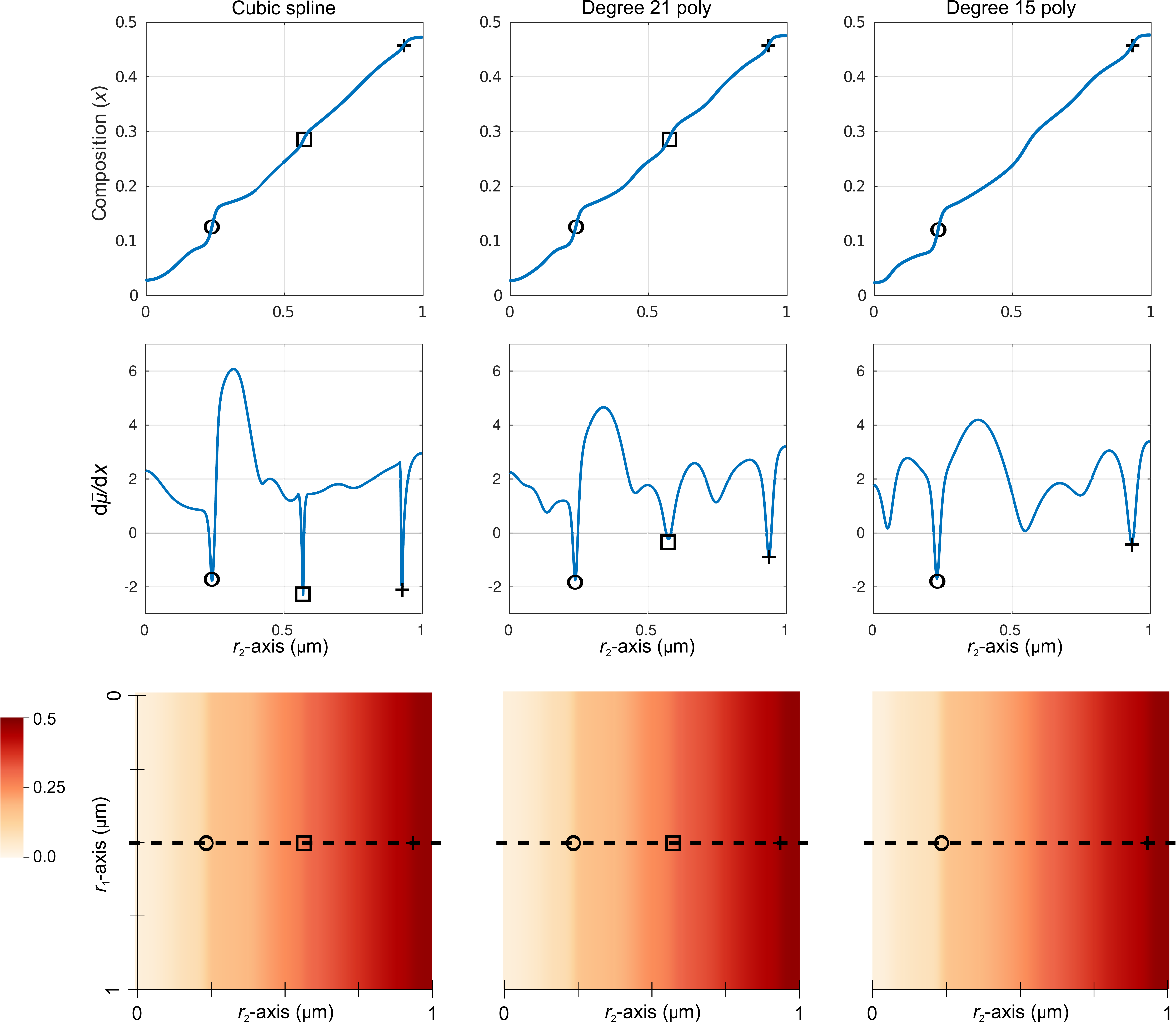}
        	\caption{Simulation results show the effect of the curve fit on the composition profile. The bottom row of plots show the 2D simulation results after 100 s, colored according to the composition. The top row of plots show the composition profile found along the dotted line in the corresponding 2D plot. The middle row shows the derivative of the homogeneous free energy with respect to composition, $\mathrm{d}\bar{\mu}/\mathrm{d}x$. The phase interfaces (spinodals) captured by the respective chemical potential curve fits occur where $\mathrm{d}\bar{\mu}/\mathrm{d}x$ is negative, and they are marked by the circle, square, and plus sign. Note that the degree 15 polynomial missed the middle spinodal, causing that interface to be smoothed out.}
	\label{fig:simResults}
\end{figure}

\subsection{Composition profiles}
To demonstrate the effect of the curve fits from the previous section, we performed 2D simulations of the diffusion of oxygen within metallic Ti at 800K. We fit the data for the excess chemical potential with the cubic spline fit shown in Figure \ref{fig:splineFit} and the R-K polynomial fits of degree 15 and 21 shown in Figure \ref{fig:polyFit}. The domain was $[0,1]\times[0,1]\, \mathrm{\upmu m}$ with a $1 \times 500$ element mesh. Initial conditions  were a uniform composition gradient between $x = 0.005$ at $r_2 = 0$ and $x = 0.495$ at $r_2 = 1$. Zero flux boundary conditions were applied. The average value of the homogeneous chemical potential is $\bar{\mu}_{avg} = -0.634$. We use $\hat{\kappa} = 2.5e-4$, which gives a length scale of about 0.02 $\mathrm{\upmu m}$. This is well resolved by the element length of 0.002 $\mathrm{\upmu m}$ in the $r_2$ direction.

Figure \ref{fig:simResults} shows the composition profile after 100 s for the three simulations, as well as the derivative of the homogeneous component of the chemical potential with respect to composition, $\mathrm{d}\bar{\mu}/\mathrm{d}x$. Recall that the two-phase regions correspond to the concave regions of the free energy, which is identified by a negative value for $\mathrm{d}\bar{\mu}/\mathrm{d}x$. While the 800K data has three spinodal regions between $x=0$ and $x=0.5$, recall from Figure \ref{fig:polyFit} that the degree 15 R-K polynomial does not capture the middle two-phase region. The effect of this is seen in Figure \ref{fig:simResults}. Note that the cubic spline and the degree 21 polynomial have a well-defined interface near $r_2 = 0.6$, but this interface is smoothed out when using the degree 15 polynomial. Additionally, because we are using a constant mobility $L$, every oscillation in the excess chemical potential function is reflected in the composition profile. Some of these oscillations are in the data itself; others result from the curve fit and can appear to be interfaces. However, even when it is difficult to identify the spinodal regions from the composition profile alone, the derivative of the homogeneous chemical potential clearly represents what is a true two-phase region.

\begin{figure}[tb!]
        \centering
        \includegraphics[width=0.6\textwidth]{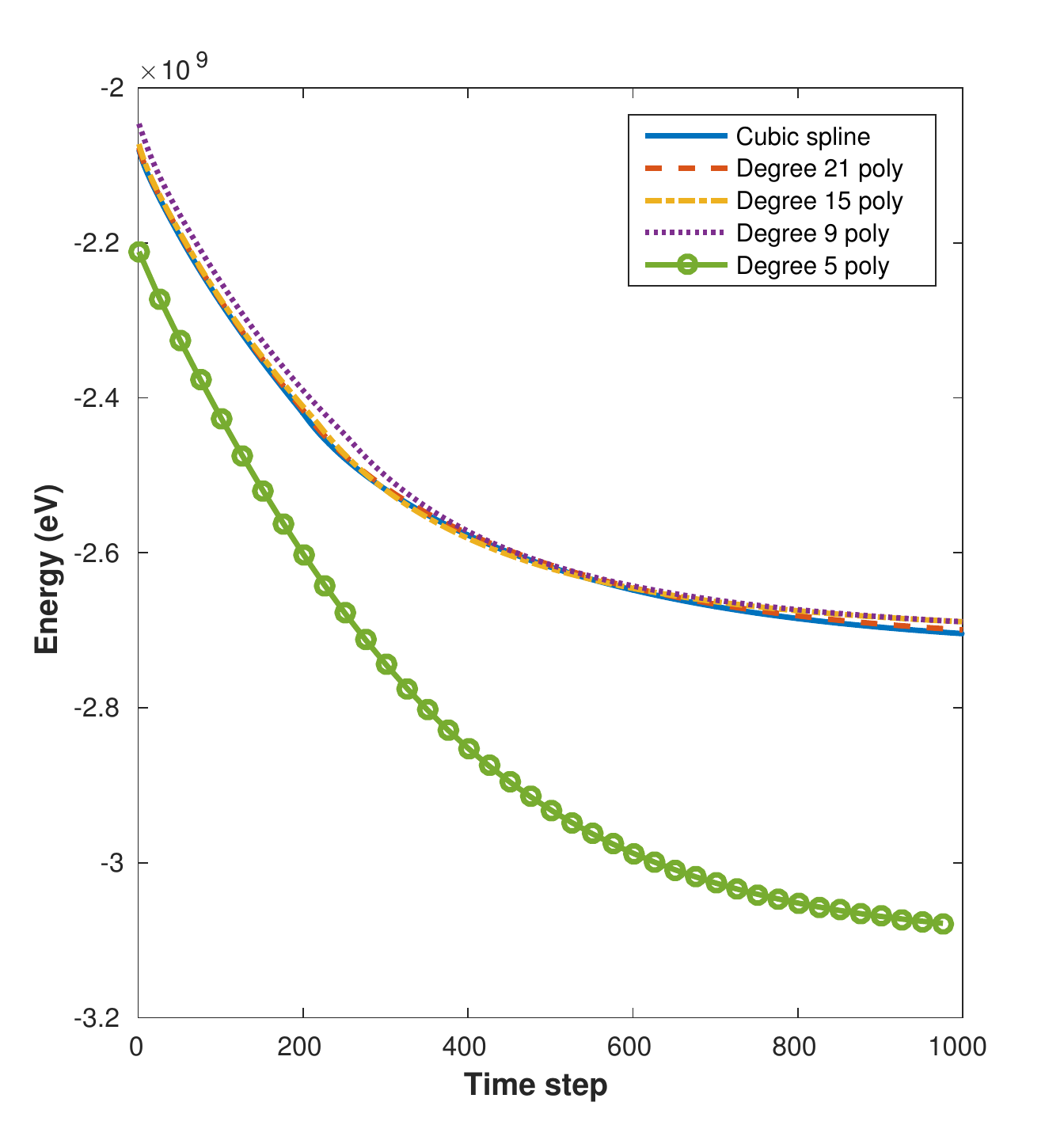}
	\caption{The time progression of total energy for a cubic spline fit is compared to polynomials of degree 5, 9, 15, and 21. Initial conditions were a uniform composition gradient of $[0,0.49]$ per micrometer over a $1 \times 500$ element mesh. The results show 1000 time steps of 10 s using a constant mobility of $9\times10^{-6} \,\mathrm{\upmu m^2/s}$.}
	\label{fig:totalEnergy}
\end{figure}

\subsection{Total free energy as a measure of accuracy}
We compare the accuracy of the free energy representations by evaluating the time evolution of their total free energies, ${\textbf F}$, given by Eq. (\ref{eqn:effective total free energy}). Two-dimensional simulations with the same conditions described previously ran for 1000 time steps of 10 s each, using a constant mobility of $9\times10^{-6}\,\mathrm{\upmu m^2/s}$. A thickness of $1 \,\mathrm{\upmu m}$ was used in the calculation of total energy. These results are compared to the simulation that used the cubic spline fit, which is considered to most accurately represent the chemical potential of the physical system (see Figure \ref{fig:totalEnergy}). We would expect a good fit of the chemical potential to compute a total energy that matches the energy when using the spline fit.

The zero flux boundary conditions, together with the initial uniform composition gradient, result in a composition profile that gradually approaches a uniform composition of $x=0.25$ in each simulation. The degree five R-K polynomial clearly overestimates the magnitude of the total free energy. To a lesser degree, the degree nine order polynomial underestimates the magnitude of the total free energy. This is consistent with the computed homogeneous free energy curves shown in Figure (\ref{fig:polyFit}), where the degree 5 polynomial fit for the excess chemical potential results in a free energy curve significantly lower than the reference curve. The degree 9 R-K polynomial is slightly above the reference. The other two R-K polynomial fits and the spline fit match the reference free energy curve relatively well, which is reflected in the simulation results of Figure \ref{fig:totalEnergy}.

\subsection{Assembly time}
We also compare the effect on computation time of using the cubic spline in comparison with the degree 21 polynomial. Evaluation of the chemical potential and its derivatives takes place when the residual and Jacobian are evaluated and assembled in the phase field code. The additional terms in the degree 21 polynomial relative to the cubic spline incur a significant increase in the number of floating point operations in the computation. To quantify this effect, 3D simulations of oxygen diffusion in metallic Ti were performed on a cube of length $2 \,\mathrm{\upmu m}$ with a $100 \times 100 \times 100$ element mesh using 160 processors. Zero flux boundary conditions were applied, with initial conditions of a uniform composition gradient between $x = 0.001$ at $r_1 = 0$ and $x = 0.499$ at $r_1 = 2$.

\begin{figure}[tb!]
        \centering
        \includegraphics[width=0.95\textwidth]{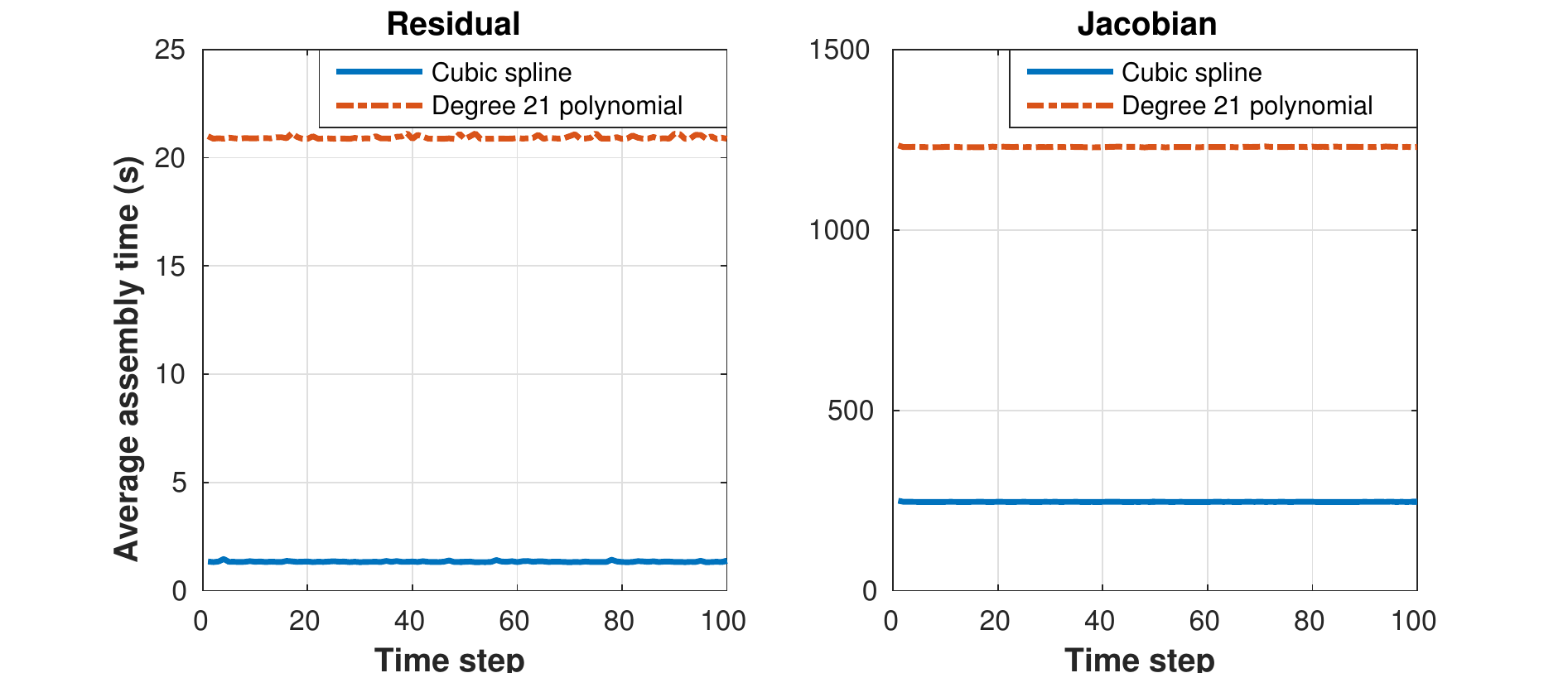}
        	\caption{The residual and Jacobian assembly wall times, averaged over each time step, for 100 time steps.}
	\label{fig:plotTime}
\end{figure}

Wall times for assembly of the residual and the Jacobian for the first time step are shown in Table \ref{tab:assemblyTime}. There is a significant difference in the assembly times, with the polynomial fit taking more that 15 times as long to evaluate and assemble the residual and about 5 times as long for the Jacobian. This difference in computation time continues long after the first time step. The wall times, averaged over each time step, are plotted in Figure \ref{fig:plotTime} for the first 100 time steps. The wall times for both simulations remain essentially constant over that time, with the residual and Jacobian taking about five and 15 times longer, respectively, when using the degree 21 polynomial instead of the cubic spline.

Note that for a system where a polynomial of lower degree provides a sufficient fit, the difference in wall time would decrease. For example, if we simulated the oxgyen diffusion in titanium at 1800K, as in Figure \ref{fig:1800fit}, a cubic polynomial would sufficiently represent the excess chemical potential, and there would be no decrease in computation time from using a cubic spline instead. However, for a complex free energy landscape, such as present in the diffusion of oxygen in Ti at 800K, the decrease in computation time achieved by using a cubic spline as opposed to a polynomial of high degree is significant.

\begin{table}[tb!]
\centering
\caption{Comparison of wall times for residual and Jacobian assembly with cubic spline and degree 21 R-K polynomial fit for the first time step.}
\begin{tabular}{c c c c c}
\hline
      Newton      &    \multicolumn{2}{c}{Residual time (s)} &   \multicolumn{2}{c}{Jacobian time (s)}\\
      \cline{2-5}
       iteration	& Polynomial & Spline & Polynomial &  Spline \\
 \hline
0 & 21.0 &	1.37 &	1230 &	249\\
1 & 21.0 &	1.40 &	1240 &	249\\
2 & 20.9 &	1.32 &	1230 &	249\\
3 & 21.3 &	1.34 &	1230 &	249\\
4 & 20.9 &	1.32 &	1240 &	250\\
\hline
\end{tabular}
\label{tab:assemblyTime}
\end{table}

\section{Conclusion}
\label{sec:concl}

Redlich-Kister polynomials have commonly been used to represent free energy functions. The global nature of the these polynomials often prevents them from capturing local phenomena. Spline functions present an effective alternative for representing chemical potential and  free energy data. Splines are piecewise polynomials with a specified order of continuity across the entire domain. Because of their piecewise structure, they are able to accurately represent local features of the data. The inherent constraints prescribing that the spline and a certain number of its derivatives be continuous across all subdomain junctions give splines an order of global continuity. These local and global properties make splines an important tool in curve fitting.

For a simple landscape, the R-K polynomial sufficiently models the chemical potential with a low polynomial degree. This is demonstrated by the high temperature data for diffusion of oxygen in metallic Ti, where a cubic R-K polynomial gave a sufficient representation of the excess chemical potential. This contrasts, however, with more complex cases involving multiple regions of phase separation, where much higher polynomial degrees are needed when using the R-K polynomials. At the lower temperature of 800K, for example, the  diffusion of oxygen in Ti produces three spinodal regions, and the R-K expansion requires a polynomial degree of around 20 to represent all three. These polynomials of high degree add complexity and can introduce spurious oscillations, but polynomials of lower degree potentially miss significant physics, such as the two-phase regions corresponding to a negative curvature of the free energy. Splines, however, are able to represent the fitted data and the associated physics with high accuracy while using a low degree.

To demonstrate the effect of the function type used to fit the excess chemical potential, we performed phase field computations of the diffusion of oxygen in titanium using a cubic spline fit and R-K polynomial fits of multiple degrees. Because the Allen-Cahn equations reach a minimum more rapidly than the Cahn-Hilliard equation, our phase field model is based on an effective Cahn-Hilliard equation. Using this model, we compared the performance of a cubic spline fit to the R-K polynomial fits in three ways. First, we observed that, for this example, R-K polynomials of degree less than 21 fail to represent all three well-defined phase interfaces, while the cubic spline does. We also compared the evolution of the total free energy of the system as it relaxes toward a homogeneous composition, and we found that the degree five and, to a lesser extent, the degree nine R-K polynomials misrepresent the magnitude of the total free energy. From these two studies, we concluded that a degree 21 R-K polynomial gives a sufficient representation of the excess chemical potential for this particular data. However, we then compared  the computation time for the assembly of the residual and Jacobian using the cubic spline and the degree 21 R-K polynomial. The computation with the cubic spline evaluated and assembled the residual with a speedup of 15$\times$ and the Jacobian with a speedup of 5$\times$ relative to the computation using the degree 21 polynomial. The ability of the cubic spline to capture all of the local physics of the system while minimizing spurious oscillations and computation time makes it a valuable tool in representing chemical potential and free energy functions.

\section*{Acknowledgements}
This work was carried out under an NSF DMREF grant: DMR1436154 ``DMREF: Integrated Computational Framework for Designing Dynamically Controlled Alloy-Oxide Heterostructures''. The computations were performed using resources supported by the U.S. Department of Energy, Office of Basic Energy Sciences, Division of Materials Sciences and Engineering under Award \#DE-SC0008637 that funds the PRedictive Integrated Structural Materials Science (PRISMS) Center at University of Michigan. 

\section*{References}
\bibliographystyle{unsrtnat}%abbrvnat}
\bibliography{references}

\end{document}